\documentclass[times,10pt,twocolumn]{article} 
\usepackage{latex8}
\usepackage{times}
\usepackage{epsfig}

\def\d3k{{\displaystyle {\rm d}{\bf k} \over \displaystyle (2\pi)^3}}
\def\hmpc{h^{-1} {\rm Mpc}}

\begin{document}

\title{Voronoi Tessellations and the Cosmic Web:\\Spatial Patterns and Clustering across the Universe}

\author{Rien van de Weygaert\\
Kapteyn Astronomical Institute, University of Groningen, the Netherlands\\ 
weygaert@astro.rug.nl\\
}

\maketitle
\thispagestyle{empty}

\begin{abstract}
The spatial cosmic matter distribution on scales of a few up to more than a hundred 
Megaparsec\footnote{The main measure of length in astronomy is the parsec. Technically 
a parsec is the distance at which we would see the distance Earth-Sun at an angle of 
1 arcsec. It is equal to 3.262 lightyears $=3.086\times 10^{13} \hbox{\rm km}$. Cosmological 
distances are substantially larger, so that a Megaparsec ($=10^6\,pc$) is the regular 
unit of distance. Usually this goes along with $h$, the cosmic expansion rate (Hubble 
parameter) $H$ in units of $100$ km/s/Mpc ($h\approx 0.71$).} displays a salient and 
pervasive foamlike pattern. Voronoi tessellations are a versatile and flexible mathematical 
model for such weblike spatial patterns. They 
would be the natural asymptotic result of an evolution in which low-density expanding void regions 
dictate the spatial organization of the Megaparsec Universe, while matter assembles in 
high-density filamentary and wall-like interstices between the voids. We describe the 
results of ongoing investigations of a variety of aspects of cosmologically 
relevant spatial distributions and statistics within the framework of Voronoi 
tessellations. Particularly enticing is the finding of a profound scaling of both 
clustering strength and clustering extent for the distribution of tessellation nodes, 
suggestive for the clustering properties of galaxy clusters. Cellular patterns may be 
the source of an intrinsic  ``geometrically biased'' clustering.
\end{abstract}

%------------------------------------------------------------------------- 
\Section{Introduction: the Cosmic Web}
Macroscopic patterns in nature are often due the collective action of basic, often 
even simple, physical processes. These may yield a surprising array of complex and 
genuinely unique physical manifestations. The macroscopic organization into complex 
spatial patterns is one of the most striking. The rich morphology of such systems and patterns 
represents a major source of information on the underlying physics. This has 
made them the subject of a major and promising area of inquiry.

One of the most striking examples of a physical system displaying a salient
geometrical morphology, and the largest in terms of sheer size, is 
the Universe as a whole. The past few decades have revealed  that on scales of a 
few up to more than a hundred Megaparsec, galaxies conglomerate into intriguing 
weblike patterns that pervade throughout the observable cosmos. Revealed through the 
painstaking efforts of redshift survey campaigns, it has completely revised our view of 
the matter distribution on these cosmological scales. 
\begin{figure}[t]
\begin{center}
\mbox{\hskip -0.5truecm\includegraphics[height=10.5cm]{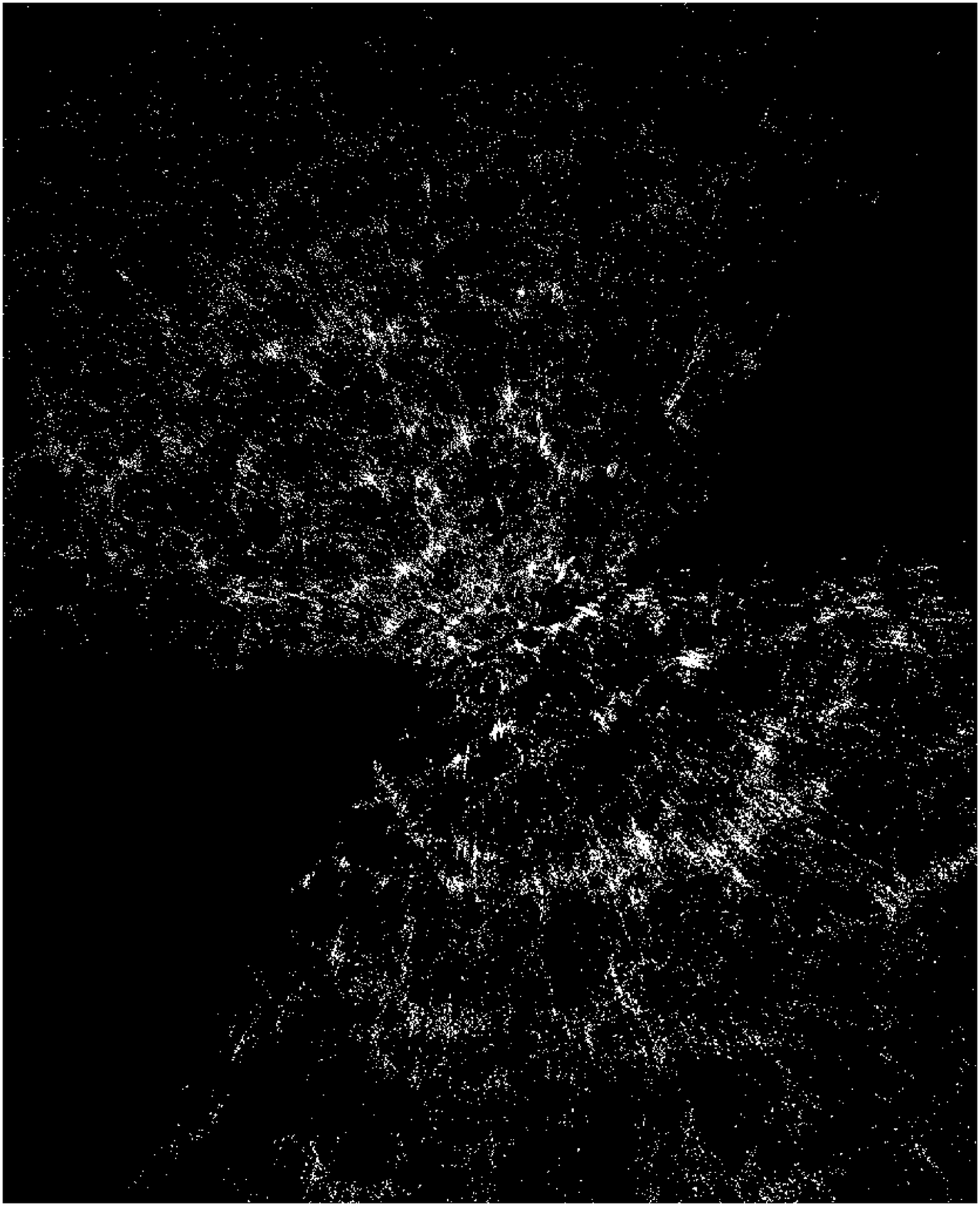}}
\vskip 0.0truecm
\caption{Image of the weblike galaxy distribution in the SDSS galaxy redshift survey (3rd data release, 
374,767 galaxies).}
\vskip -1.0truecm
\end{center}
\label{fig:sdssgaldist}
\end{figure}
The spatial distribution of galaxies is far from homogeneous. Instead, we recognize a weblike 
arrangement. Galaxies aggregate in striking geometric patterns, outlined by huge {\it filamentary} 
and {\it sheetlike} structures, the sizes of the most conspicuous ones frequently exceeding 100$\hmpc$. 
Within and around these anisotropic features we find a variety of density condensations, ranging from modest 
groups of a few galaxies up to massive compact {\it clusters} of galaxies. The latter represent the 
most prominent density enhancements in our universe and usually mark the dense intersections of 
filaments and sheets. They stand out as the most massive and most recently formed (individual) objects in 
the Universe. Complementing this cosmic inventory leads to the existence of large {\it voids}, enormous 
regions with sizes in the range of $20-50\hmpc$ that are practically devoid of any galaxy, usually roundish 
in shape and occupying the major share of space in the Universe. 

Of utmost significance for our inquiry into the issue of cosmic structure formation is 
the fact that the prominent structural components of the galaxy distribution -- clusters, 
filaments, walls and voids -- are not merely randomly and independently scattered features. 
On the contrary, they have arranged themselves in a seemingly highly organized and structured fashion, 
woven into an intriguing {\it weblike} tapestry that permeates the whole of the explored Universe.
The weblike spatial arrangement of galaxies and mass into elongated filaments, sheetlike walls and dense compact 
clusters, the existence of large near-empty void regions and the hierarchical nature of this mass 
distribution -- marked by substructure over a wide range of scales and densities -- are three 
major characteristics of what we have come to know as the {\it Cosmic Web}. Its appearance is most dramatically 
illustrated by the most recently produced maps of the nearby cosmos. The 2dF -- two-degree field -- Galaxy 
Redshift Survey (e.g. \cite{colless2003}) mapped the spatial distribution of nearly 250,000 galaxies in two 
narrow sections through the local Universe, out to a depth of more than 300$\hmpc$. The SDSS survey (fig.~\ref{fig:sdssgaldist}, 
see e.g.~\cite{tegmark2004}), is mapping the location of up to a million galaxies, covering nearly a quarter of the 
sky and reaching out to a depth of beyond 500$\hmpc$. 

\Section{Voronoi Models \& the Cosmic Web}
In the cosmological context {\it Voronoi Tessellations} represent the {\it 
Asymptotic Frame} for the ultimate matter distribution distribution in any 
cosmic structure formation scenario. The Voronoi tessellation is the {\it skeleton}  
of the cosmic matter distribution, identifying the structural frame around which 
matter will gradually assemble during the emergence of cosmic structure.

Voronoi tessellations are a versatile and flexible mathematical model for foamlike spatial patterns. 
They would be the natural result of an evolution in which expanding {\it voids} dictate the spatial organization 
of the Megaparsec Universe, with matter assembling in the high-density filamentary 
and wall-like interstices between the voids. According to recent work this is indeed what may be expected in 
standard scenarios of cosmic structure formation. Voronoi models would represent the asymptotic limit for which 
the population of voids would correspond to one single void size and excess expansion rate. In this paper 
we seek to sketch the background and ramifications of this idea. 

The premise is that some primordial cosmic process generated a random (Gaussian) 
density fluctuation field. The troughs, minima, in this field will become 
the centres of expanding voids. Matter will flow away until it runs into its 
surroundings and encounters similar material flowing out of adjacent voids. The 
spatial distribution of the density troughs in the primordial density field 
is dependent on the specific cosmological structure formation scenario at hand. 

\begin{figure*}[t]
\vskip -12.5truecm
\begin{center}
\mbox{\hskip -0.5truecm\includegraphics[width=18.0cm]{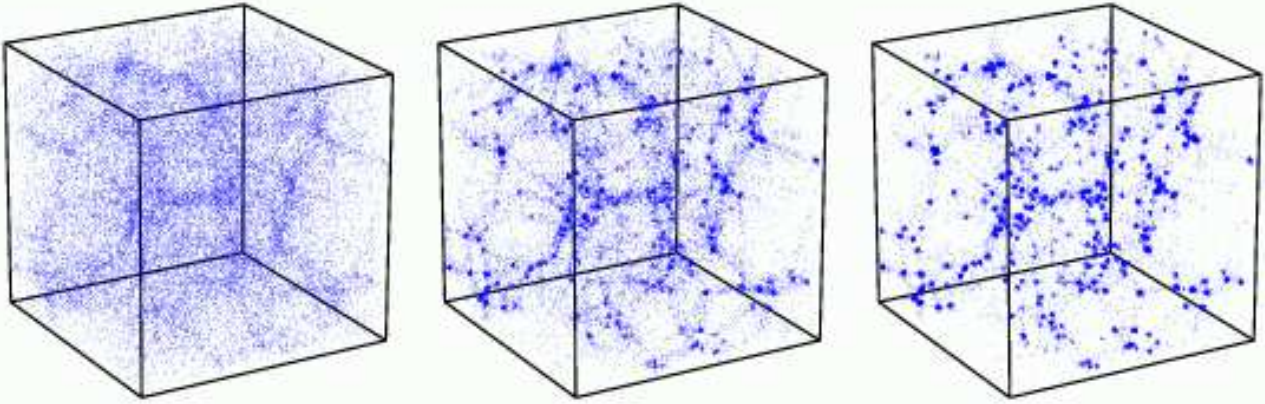}}
\vskip -0.5truecm
\caption{A sequel of three consecutive timesteps within the kinematic Voronoi cell 
formation process. The depicted boxes  have a size of $100h^{-1} \hbox{Mpc}$. Within 
these cubic volumes some $64$ Voronoi cells with a typical size of $25h^{-1}\hbox{Mpc}$ 
delineate the cosmic framework with 32000 galaxies.}
\vskip -1.0truecm
\label{fig:vorkinmcube3}
\end{center}
\end{figure*}

Within the cellular Voronoi skeleton the interior of the Voronoi {\it cells} corresponds 
to voids. The Voronoi {\it planes} are identified with walls of galaxies. 
The {\it edges} delineating the rim of each wall are identified with the filaments in the galaxy
distribution.  The most outstanding structural elements are the {\it
vertices}, corresponding to the very dense compact nodes within the cosmic web, the rich 
clusters of galaxies. In general, what is denoted as a flattened {\it supercluster} will consist of 
an assembly of various connecting walls in the Voronoi foam. The elongated {\it superclusters} or {\it filaments} 
usually consist of a few connected edges. This may be clearly appreciated from Fig.~\ref{fig:vorkinmcube3}.

\SubSection{Voronoi Virtues}
Cosmologically, the great virtue of the {\it Voronoi foam} is that it provides a conceptually 
simple model for a cellular or foamlike distribution of galaxies. By using these geometrically constructed 
models one is not restricted by the resolution or number of particles. A cellular structure can be generated 
over a part of space beyond the  reach of any N-body experiment. Even though the model does not and cannot addres
the galaxy distribution on small scales, it is nevertheless a useful prescription for the spatial 
distribution of the walls and filaments themselves. In all, it makes the Voronoi model particularly suited
for studying the properties of galaxy clustering in spatial cellular patterns. Its ease and versatility of 
construction, and its flexibility with respect to defining cosmological parameters, has made it into an ideal tool 
for statistical studies and tests of structure finding and identification techniques. 
\begin{figure}[t]
\vskip -0.0truecm
\includegraphics[width=8.5cm]{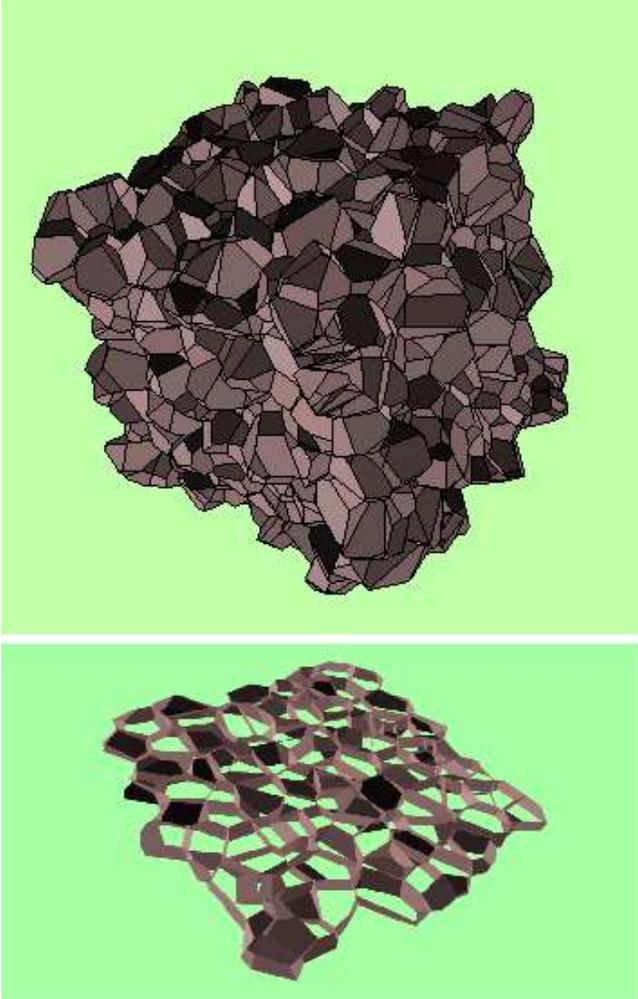}
\vskip -0.0truecm
\caption{A full 3-D tessellation comprising 1000 Voronoi cells/polyhedra 
generated by 1000 Poissonian distributed nuclei. Courtesy: Jacco Dankers}
\label{fig:vortess}
\end{figure}
\Section{Gravitational Instability}
Comprising features on a typical scale of tens of Megaparsec, the cosmic web offers a 
direct link to the matter distribution in the primordial Universe and contains a wealth 
of information on the cosmic structure formation process\footnote{According to our 
latest insights the Universe is $13.7$Gyr old. With the COBE and WMAP microwave background 
telescopes we have observed tiny disturbances in the early Universe, when the Universe 
was around 379.000 yrs old. These perturbations were no larger than a factor $10^{-5}$.}. 
It thus represents a key to unravelling one of the most pressing enigmas in modern astrophysics, 
the rise of the wealth and variety of structure in the present-day Universe from an almost 
perfectly smooth, virtually featureless, pristine cosmos.

The generally accepted theoretical framework for the formation of structure is that of 
{\it gravitational instability}. The formation and moulding of structure is ascribed to the 
gravitational growth of tiny initial density- and velocity deviations from the global cosmic 
density and expansion. Overdense regions will initially expand slightly less rapid than the cosmic 
background, reach a maximum size, turn around and ultimately condense -- dependent on 
the scale of the density perturbation -- into a recognizable astrophysical object. Underdense regions, 
on the other hand, do have a gravity deficit with respect to the surrounding Universe.  

\begin{figure}[t]
     \vskip -0.0cm
  \begin{center}
  \mbox{\hskip -0.0truecm\includegraphics[width=7.0cm]{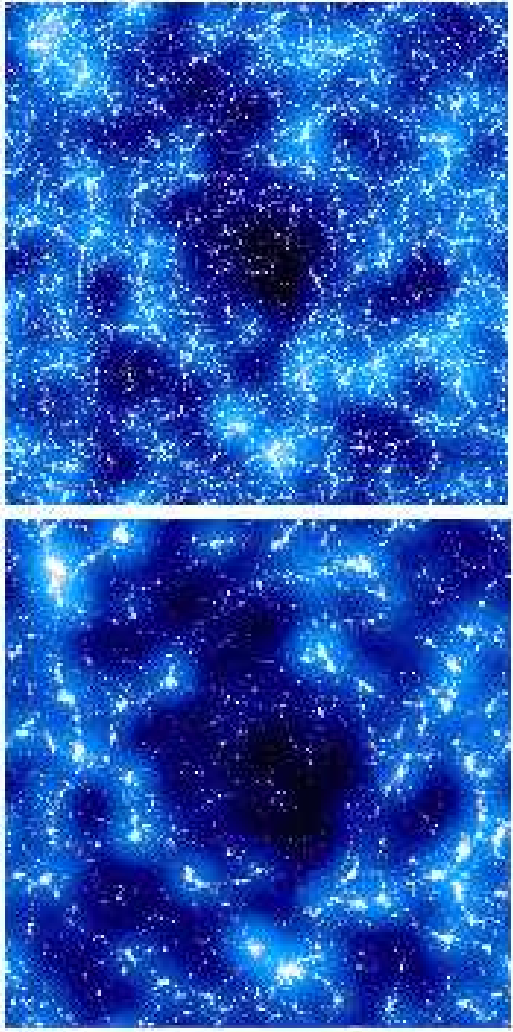}}
  \end{center}
    \vskip -0.5truecm
\caption{Void evolution: two timesteps in the evolution of a void region in a N-body 
computer simulation of structure formation (in a SCDM model). Courtesy: Erwin Platen.}
\label{fig:voidhierevol}
\vskip -0.75truecm
\end{figure}

Three fundamental aspects of the ensuing nonlinear gravitational clustering process determine the 
morphology of the resulting matter distribution. The {\it first} is {\it hierarchical} 
clustering: the first objects to form are small compact objects which subsequently merge with their 
surroundings into ever larger features. The {\it second} fundamental aspect concerns {\it anisotropic 
gravitational collapse}. Aspherical overdensities, on any scale and in any scenario, will contract such 
that they become increasingly anisotropic. At first they turn into a flattened `pancake', rapidly followed 
by contraction into an elongated filament, possibly to finally collapse to become a galaxy or a cluster.
The {\it third} manifest feature of the Megaparsec Universe is the marked and dominant presence of large roundish 
underdense regions, the {\it voids} (see fig.~\ref{fig:voidhierevol}). 

\SubSection{Voids}
Inspired by early computer calculations, \cite{icke1984} pointed out that for the 
understanding of the formation of the large coherent patterns pervading the 
Universe it is worthwhile to direct attention to the complementary evolution of underdense 
regions. The gravity deficit in their interior makes them expand with respect to the 
background Universe. Meanwhile we see the matter assembling in ever dense {\it planar} 
and/or {\it filamentary} interstices. 

They form in and around density troughs in the primordial density field. Because of their lower interior gravity 
they will expand faster than the rest of the Universe, while their internal matter density rapidly decreases as 
matter evacuates their interior (see fig.~\ref{fig:voidhierevol}). They evolve in the nearly 
empty void regions with sharply defined boundaries marked by filaments and walls. Their essential role in the 
organization of the cosmic matter distribution got recognized early after their discovery. 
Recently, their emergence and evolution has been explained within the context of hierarchical gravitational scenarios 
\cite{shethwey2004}. 

As the voids expand they start to take up an increasingly major fraction of the cosmic volume. \cite{icke1984} made 
the additional interesting observation that the outward expansion of voids is accompanied by a tendency of voids to 
assume a spherical geometry. While the above properties refer to the evolution of isolated voids, we know that 
in reality they will be surrounded by other density structures and that they 
will themselves contain a range of smaller scale features and objects. Recent work by 
\cite{shethwey2004} managed to describe the corresponding {\it ``Void Sociology''} through an 
analytical description within the context of the excursion set formalism \cite{bond1991}. 
Fig.~\ref{fig:voidhierevol} contains an illustration of the evolving Cosmic Web, marked by 
the surge and development of large cosmic voids. The evolution of voids is dictated 
by two processes:

\noindent $\bullet$ Voids merge with surrounding voids into a larger void.

\noindent $\bullet$ Voids may get destroyed when embedded within a larger collapsing overdensity.

\medskip
\noindent When working out the implication of the precise interplay between these two processes, one 
finds that voids tend to define a population marked by a self-similar peaked size distribution centered 
around a characteristic void size: at any one cosmic instant the void population is dominated by voids of 
nearly the same size (fig.~\ref{fig:voidhierevol}). This indeed forms a telling 
confirmation of the observed foamlike distribution of galaxies with the characteristic pattern 
punctuated by voids whose size is $\approx 10-30h^{-1}\ \hbox{\rm Mpc}$. 
\SubSection{From Voids to Voronoi}
\label{sec:voidvoronoi}
A bold leap leads to a geometrically interesting situation. Taking the voids as the dominant dynamical 
component of the Universe, we may think of the Megaparsec  scale structure as a close packing of spherically 
expanding regions. The asymptotic configuration for a ``peaked'' void distribution is one of a single 
characteristic void size. Pursuing this situation we would find a cosmic matter distribution which would be 
organized by a population of {\em equally sized} and {\it equally fast expanding}, {\it spherically shaped} voids. 

Matter will collect at the interstices between the expanding voids. In this asymptotic limit of the outflow being 
the same in all voids, these interstices are the {\it bisecting planes} between two neighbouring expansion centres: 
the {\it walls} and {\it filaments} would be found precisely at the (bisecting) midplanes between expanding voids. 
For any given set of expansion centres, or {\it nuclei}, the arrangement of these planes defines a unique process 
for the partitioning of space: the resulting skeleton of the matter distribution would be nothing else than 
a {\it Voronoi tessellation} \cite{vor1908,okabe2000} (see fig.~\ref{fig:vortess}). In other words, 
the {\it ASYMPTOTIC} description of the cosmic clustering process leads to a geometrical configuration 
that is one of the main concepts in the field of stochastic geometry:\\
\ \\
\centerline{{\Large VORONOI TESSELLATIONS}}
\ \\
A cosmological realisation of this process is called a {\it Voronoi foam} \cite{ickewey1987}.

%------------------------------------------------------------------------- 
\Section{Voronoi Clustering Models}
{\it Voronoi Clustering Models} are a class of heuristic models for
cellular distributions of matter which use the Voronoi tessellation as the skeleton of the 
cosmic matter distribution \cite{weyicke1989,weygaert1991,weygaert2007a}. 

It is the stochastic yet non-Poissonian geometrical distribution of the {\it walls}, 
{\it filaments} and {\it clusters} embedded in the cosmic web which generates 
the large-scale clustering properties of matter and the related galaxy populations. 
It is precisely this aspect which is modelled in great detail by Voronoi tessellation. 
The small-scale distribution of galaxies, i.e. the distribution within the various 
components of the cosmic skeleton, will involve the complicated details of highly 
nonlinear small-scale interactions of the gravitating matter. Ideally, well-defined 
and elaborate physical models and/or N-body computer simulations would fill in this 
aspect. In the Voronoi models described here we complement the geometrically 
fixed configuration of the Voronoi tessellations with a heuristic prescription for 
the placing of particles/model galaxies within the tessellation. 

\begin{figure}[t]
  \centering
  \vskip 0.0truecm
  \mbox{\hskip 0.0truecm\includegraphics[height=7.0cm]{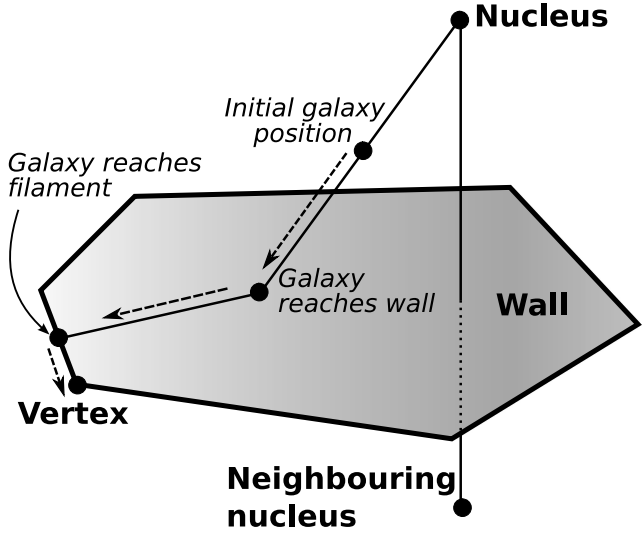}}
  \vskip 0.0cm
  \caption{Schematic illustration of the Voronoi kinematic model. Courtesy: Miguel Arag\'on-Calvo.}
  \label{fig:vorkinmschm}
  \vskip -0.75cm
\end{figure} 
The {\it Voronoi Kinematic Voronoi Model} is based upon the notion
that voids play a key organizational role in the development of
structure and make the Universe resemble a soapsud of expanding
bubbles \cite{icke1984}. It forms an idealized and asymptotic description
of the outcome of the cosmic structure formation process within
gravitational instability scenarios with voids forming around a dip in
the primordial density field. This is translated into a scheme for the 
displacement of initially randomly distributed galaxies within the Voronoi 
skeleton. Within a void, the mean distance between galaxies increases uniformly 
in the course of time. When a galaxy tries to enter an adjacent cell, the
velocity component perpendicular to the cell wall disappears. Thereafter, the galaxy 
continues to move within the wall, until it tries to enter the next cell; it then loses 
its velocity component towards that cell, so that the galaxy continues along a
filament. Finally, it comes to rest in a node, as soon as it tries to
enter a fourth neighbouring void.

\SubSection{Voronoi Kinematic Model}
\noindent All different Voronoi models are based upon the displacement
of a sample of $N$ {\it Voronoi galaxies}. The initial spatial
distribution of these $N$ galaxies within the sample volume $V$ is
purely random, their initial locations ${\bf x}_{n0}$ defined by a 
homogeneous Poisson process. A set of $M$ nuclei or {\it
expansion centres} within the volume $V$ corresponds to the cell
centres, or {\it expansion centres} driving the evolving matter
distribution.
 
\bigskip
\noindent The first stage of the procedure consists of the generation of initial conditions of the Voronoi galaxy 
distribution, along with the specification of the properties (width) of the structural elements:

\medskip
\noindent $\bullet$ Distribution of $M$ nuclei, {\it expansion centres}, within the simulation volume $V$. The location 
of nucleus $m$ is ${\bf y}_m$. 

\noindent $\bullet$ Generate $N$ model galaxies whose initial locations, ${\bf x}_{n0}$ ($n=1,\ldots,N)$ 
are randomly distributed throughout the sample volume $V$.

\noindent $\bullet$ Of each model galaxy $n$ determine the Voronoi cell ${\cal V}_{\alpha}$ in which it is located, ie. 
determine the closest nucleus $j_{\alpha}$.

\noindent $\bullet$ The width of the walls, filaments and vertices is set by assuming them to have a Gaussian radial 
density distribution, with wall width $R_{\rm W}$, filament width $R_{\rm F}$ and vertex width $R_{\rm V}$ input 
parameters of the Voronoi model.
%\begin{eqnarray}
%\rho_{\rm ft}(r)&\,=\,&{1 \over ({2 \pi} R_{\rm ft}^2)^{1/2}}\,\exp \left(- {r_{\rm ft}^2 \over 2 R_{\rm ft}^2}\right)\,.
%\end{eqnarray}

\bigskip
\noindent The second stage of the procedure consists of the calculation of the complete {\it Voronoi track} for each galaxy $n=1,\ldots,N$. 
Figure~\ref{fig:vorkinmschm} contains a sketch of a typical Voronoi galaxy track: 

\medskip
\noindent $\bullet$ The first step is the calculation of the galaxy tracks is the 
dtermination for each galaxy $n$ the Voronoi cell ${\cal V}_{\alpha}$ in which 
it is initially located, ie. finding the nucleus $j_{\alpha}$ which is closest 
to the galaxies' initial position ${\bf x}_{n0}$.

\noindent $\bullet$ Subsequently, the galaxy follows a radial path within Voronoi cell ${\cal V}_{\alpha}$. 
It moves along the path emanating from its expansion centre $j_\alpha$, ie. along the direction marked 
by the unity vector ${\hat {\bf e}}_{n\alpha}$ defined by the line starting at ${\bf y}_{\alpha}$ and 
pointing radially outward from $j_{\alpha}$, ${\bf r}_{n\alpha}\,\equiv\,{\bf y}_{\alpha}\,+\,R_n\,{\hat {\bf e}}_{n\alpha}$. 
This is pursued until the galaxy intersects the Voronoi wall $\Sigma_{\alpha\beta}$. 

\noindent $\bullet$ After this the galaxy's displacement is restricted to the wall $\Sigma_{\alpha\beta}$ as 
the displacement component perpendicular to the wall is damped. The galaxy moves towards the 
one edge $\Lambda_{\alpha\beta\gamma}$ defining the nearest intersection with wall path 
$r_{n\alpha\beta}$. This continues until it intersects Voronoi edge 
$\Lambda_{\alpha\beta\gamma}$. 

\noindent $\bullet$ Subsequently, it pursues its path along the spine of the edge $\Lambda_{\alpha\beta\gamma}$ as 
the velocities' component perpendicular to the edge has been suppressed (``the self-gravity of the 
filament damps its corresponding velocity''). This continues till the galaxy finally arrives 
at vertex $\Xi_{\alpha\beta\gamma\delta}$. The vertex is the final destination of the galaxy, 
the deep potential well of the cluster at its location will keep the galaxy within its 
reach. 

\bigskip
By determining the total {\it Voronoi track} for each galaxy $n$ it is rather straightforward 
to compute the location of the galaxy at any cosmic epoch $t$ by determining the {\it
displacement} ${\bf x}_n$ that each galaxy has traversed along its path. For each 
cosmic epoch we may therefore easily compute the corresponding mass distribution. 
\begin{figure*}[t]
\begin{center}
\includegraphics[width=16.5cm]{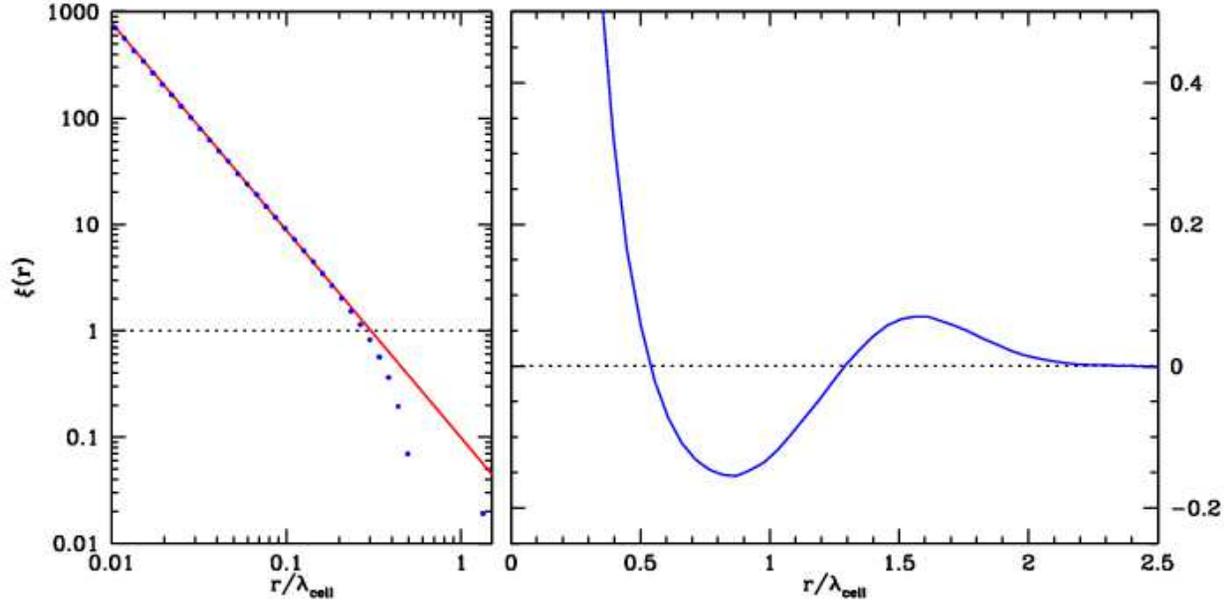}
\end{center}
\vskip -0.75truecm
\caption{Two-point correlation function of (Poisson) Voronoi vertices. Distance in units of 
inter-nucleus distance. Left: log-log plot of $\xi(r)$. Right: lin-lin plot of $\xi(r)$.}
\label{fig:vorvrtxi}
\vskip -0.75truecm
\end{figure*}
\SubSection{Kinematic Evolution}
The evolutionary progression within the Voronoi kinematic scheme, from an almost 
featureless random distribution, via a wall-like and filamentary morphology towards 
a distribution in which matter ultimately aggregates into conspicuous compact 
cluster-like clumps can be readily appreciated from the sequence of 6 cubic 3-D 
particle distributions in Figure~\ref{fig:vorkinmcube3} \cite{weygaert2007a}. 

The steadily increasing contrast of the various structural features is accompanied by a 
gradual shift in topological nature of the distribution. The virtually uniform particle 
distribution at the beginning (upper lefthand frame) ultimately unfolds into the 
highly clumped distribution in the lower righthand frame. At first only a faint 
imprint of density enhancements and depressions can be discerned. In the subsequent 
first stage of nonlinear evolution we see a development of the matter distribution 
towards a wall-dominated foam. The contrast of the walls with respect to the general 
field population is rather moderate (see e.g. second frame), and most obviously 
discernable by tracing the sites where the walls intersect and the galaxy density is 
slightly enhanced. The ensuing frames depict the gradual progression via a wall-like 
through a filamentary towards an ultimate cluster-dominated matter distribution. By 
then nearly all matter has streamed into the nodal sites of the cellular network. The 
initially almost hesitant rise of the clusters quickly turns into a strong and 
incessant growth towards their appearance as dense and compact features which 
ultimately stand out as the sole dominating element in the cosmic matter distribution 
(bottom righthand frame). 

\Section{Superclustering vs. Vertex Clustering}
Upon quantitatively testing the spatial distribution of galaxies various 
interesting properties of clustering within Voronoi foams are revealed. The most 
salient and outstanding properties concern the distribution of the Voronoi 
vertices. A first inspection of the spatial distribution of Voronoi vertices 
(Fig.~\ref{fig:vorvrtxi}, bottom righthand frame) immediately reveals that it is not a 
simple random Poisson distribution. 

The impression of strong clustering, on scales smaller than or of the order of 
the cellsize $\lambda_{\rm c}$\footnote{in the following we use the cellsize $\lambda_c$, ie. the  
intranucleus distance, as unit of distance for the generated Voronoi foams.}, is most evidently expressed 
by the corresponding two-point correlation function $\xi(r)$ (Fig.~\ref{fig:vorvrtxi}). Not only one 
finds a clear positive signal out to a distance of at least $r \approx 1/4\,\lambda_{\rm c}$ but 
also -- surprising at the time of its finding on the basis of similar computer 
experiments~\cite{weyicke1989} -- the correlation function appears to be an almost perfect 
power-law, 
\begin{eqnarray}
\xi_{\rm vv}(r)&\,\equiv\,&{\displaystyle \langle n_v({\bf x}+{\bf r})\,n_v({\bf x})\rangle \over \displaystyle {\bar n_v}^2}\,-\,1\,\approx\,
\left({\displaystyle r_{\rm o} \over \displaystyle r}\right)^\gamma\,,\nonumber\\
\gamma&\,=\,&1.95;\quad r_{\rm o} \approx 0.3\,\lambda_{\rm c}\,.
\end{eqnarray}
The solid line in the log-log diagram in Fig.~\ref{fig:vorvrtxi} represents the power-law with 
these parameters, $\gamma \approx 1.95$ and $r_{\rm o} \approx 0.3\,\lambda_{\rm c}$. 

Beyond this range, the power-law behaviour breaks down and, following a
gradual decline, the correlation function rapidly falls off to a zero value once
distances are of the order of (half) the cellsize. Assessing the behaviour of $\xi(r)$ 
in a linear-linear plot, we get a better idea of its behaviour around the 
zeropoint ``correlation length'' $r_{\rm a}\approx 0.5\lambda_c$ (bottom righthand frame 
fig.~\ref{fig:vorvrtxi}). Beyond $r_{\rm a}$ the distribution of Voronoi vertices
is practically uniform. Its only noteworthy behaviour is the gradually declining and 
alternating quasi-periodic ringing between positive and negative values similar to that 
we also recognized in the ``galaxy'' distribution, a vague echo of the
cellular patterns which the vertices trace out. Finally, beyond $r \approx 2 \lambda_c$ 
any noticeable correlation seems to be absent. 

The correlation function of Voronoi vertices is a surprisingly good and solid match to the observed 
world. Maps of the spatial distribution of clusters of galaxies show that clusters themsvelves are 
not Poissonian distributed. On the contrary, they are highly clustered and 
aggregrate into huge supercluster complexes \cite{oort1983,bhc1988}. Such superclusters represent 
moderate density enhancements on a scale of tens of Megaparsec, still co-expanding with Hubble 
flow, be it at a slightly decelerated rate. 

It sheds an alternative view on the power-law clustering with power law $\gamma \approx 2$ found in 
the cluster distribution. Also, the observed cluster clustering length $r_{\rm o} \approx 
20h^{-1}\,\hbox{Mpc}$ can be explained within the context of a cellular model, suggesting a cellsize 
of $\lambda_{\rm c} \approx 70h^{-1}\,\hbox{Mpc}$ as the basic scale of the cosmic foam. 
\begin{figure*}
\begin{center}
\vskip -0.5truecm
\mbox{\hskip -0.7truecm\includegraphics[width=16.25cm]{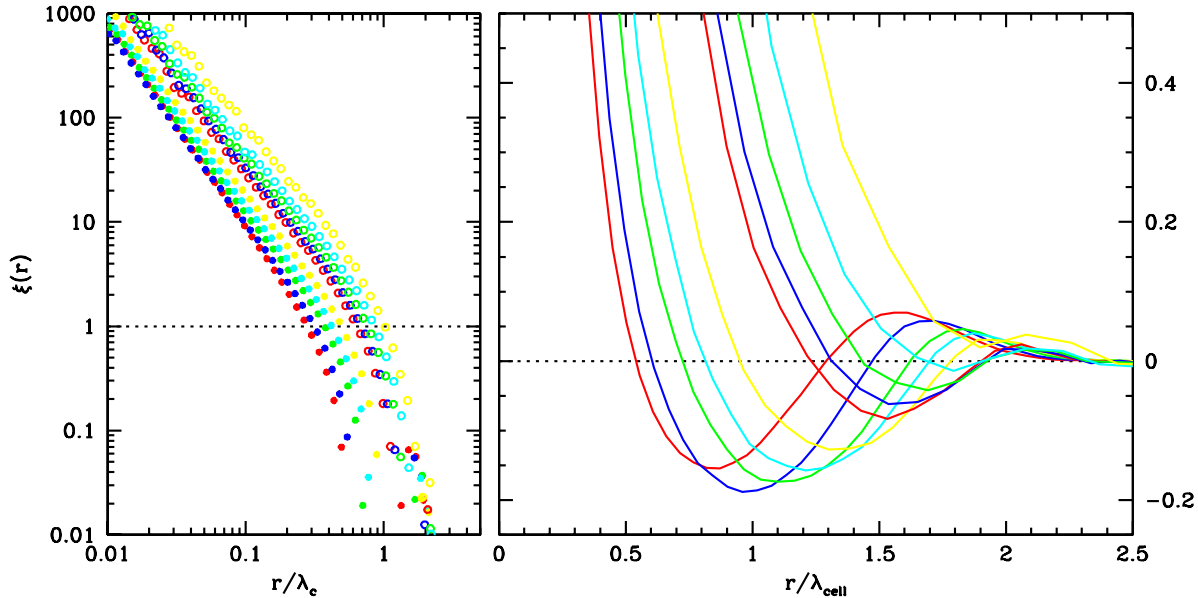}}
\vskip -0.5truecm
\end{center}
\caption{The two-point correlation for a variety of vertex subsamples, selected on the basis of ``richness/mass''. 
The largest subsample (with weakest correlation function) contains all vertices, the riches only the 2.5\% most 
massive ones. Left: log-log plot of $\xi(r)$. Right: lin-lin plot of $\xi(r)$.}
\label{fig:xicorrvrtxscale}
\vskip -0.5truecm
\end{figure*}
On the other hand, this also reveals a complication. The suggested cell scale seems to be 
well in excess of the $25h^{-1}-35h^{-1}\,\hbox{Mpc}$ size of the voids in the galaxy distribution. 
Also, it does not correspond to the clustering of objects in the walls and filaments of the 
same tessellation network when tying it to the observed galaxy-galaxy correlation. The solution to 
this dilemma leads to an intriguing finding for the scaling of vertex clustering.
\Section{Geometric Scaling} 
A {\it prominent} characteristic of superclustering in the observational world 
is that clustering of clusters is considerably more pronounced than that of 
galaxies. This characteristic ties in with a generally observed property of 
clustering to be dependent on the nature of objects. Amongst clusters 
we observe a trend for more luminous/massive clusters to be more strongly 
clustered. While the two-point correlation function $\xi_{\rm cc}(r)$ 
of clusters is consistent with it being a scaled version of the power-law galaxy-galaxy 
correlation function, with almost the same slope $\gamma \approx 1.8$, the correlation 
amplitude and clustering length $r_0$ increase as the characteristic ``mass'' of 
the clusters in the sample gets larger.
\begin{figure*}[t]
\begin{center}
\hskip -0.5truecm
\includegraphics[width=14.5cm]{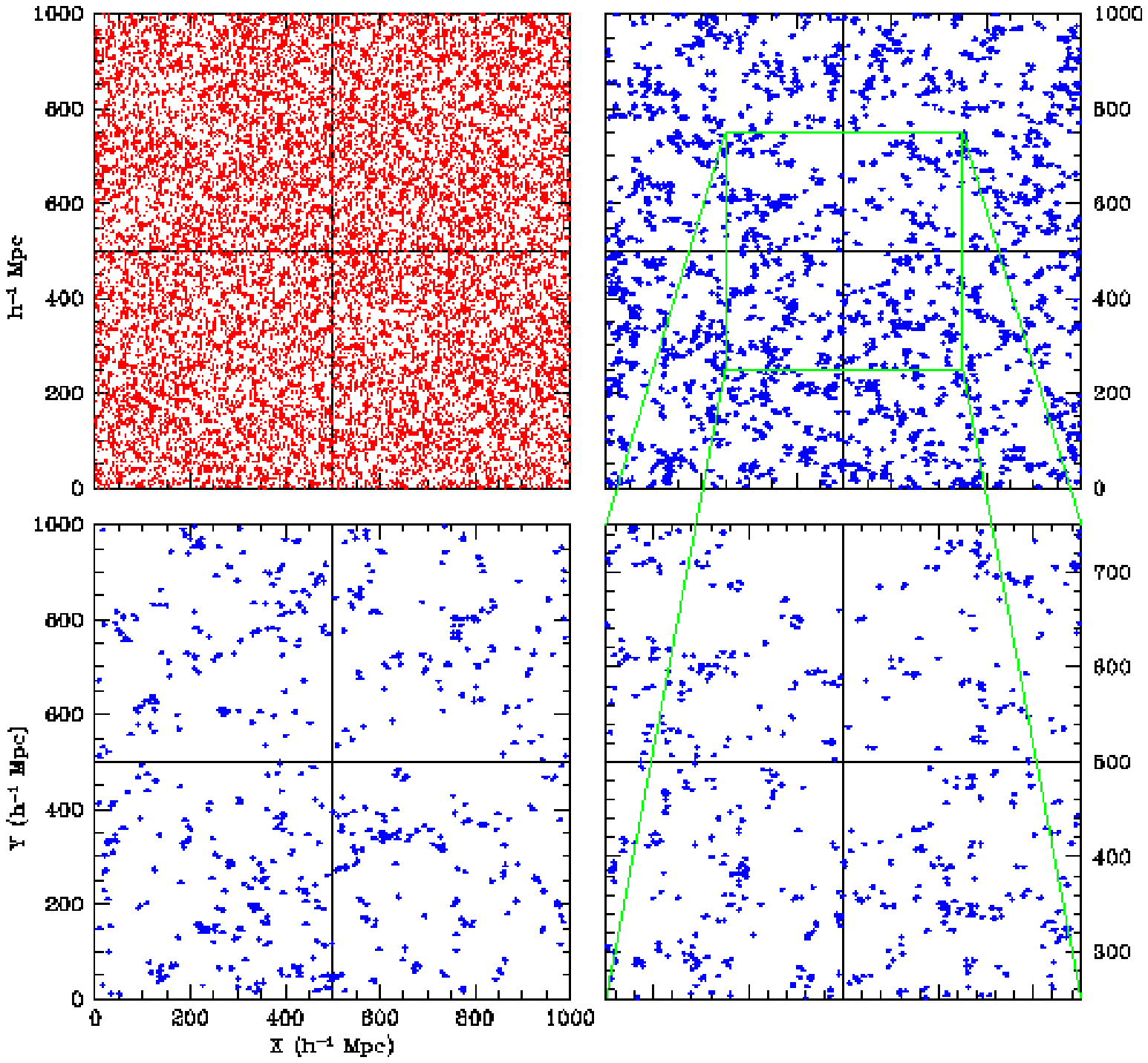}
\vskip -0.5truecm
\caption{Illustration of the `self-similar' vertex distribution. The more massive vertex 
samples are (1) more strongly clustered and (2) define clustering patterns on a larger spatial 
scale. When properly scaled a pure self-similar pattern emerges. See text for explanation.}
\end{center}
\vskip -0.3truecm
\label{fig:vorvrtxselfsim}
\end{figure*}
\SubSection{Vertex Selections}
The kinematic Voronoi model allows a geometric modelling of mass-dependent 
clustering properties by assigning a ``mass'' to each vertex. Brushing crudely over the 
details of the temporal evolution, we may assign each Voronoi vertex a ``mass'' by equating 
this to the total amount of matter which according to the ``Voronoi streaming'' description 
will ultimately flow towards that vertex. The related nuclei are the ones that supply the 
Voronoi vertex with inflowing matter. Evidently, vertices surrounding large cells 
are expected to be more massive. It is reasonably straightforward if cumbersome to calculate the  
``vertex mass'' ${\cal M}_{\rm V}$ by pure geometric means. The details, turn out to be 
challengingly complex, as it concerns the (purely geometric) calculation of the volume of a 
non-convex polyhedron centered on the Voronoi vertex.

In our computer experiments~\cite{weygaert2007a,weygaert2007b} we set up realizations of a (Poisson) 
Voronoi foam comprising 1000 cells with an average size of $25h^{-1}\hbox{Mpc}$. From the full 
vertex distribution we selected subsets of vertices, each subset comprising vertices with a 
progressively higher lower mass limit. 

A telling illustration of the significant stronger clustering for more massive vertices is 
shown in Fig~\ref{fig:vorvrtxselfsim}. It shows the vertex distribution in 
a central slice through the full 3-D cubic distribution. While the top lefthand panel 
shows the distribution of the full Voronoi vertex distribution, the distribution of the 
12.5$\%$ richest vertices (top righthand panel) and the 2.5$\%$ richest vertices (lower lefthand 
panel) is obviously significantly different. Not only do we observe a marked increase 
in clustering strength as the sample includes more massive vertices, also the spatial 
extent of the clustering patterns appears to grow as a function of vertex richness. 
When correcting for the possibly confusing influence of the sampling dilution, by sampling an 
equal number of vertices from each ``selected'' sample, the effect is even more prominent. When 
lifting the central 1/8$^{th}$ region out of the $20\%$ vertex subsample in the (top righthand) 
frame and sizing it up to the same scale as the full box, we observe the similarity in point 
process between the resulting (bottom righthand) distribution and that of the $2.5\%$ subsample 
(bottom lefthand). It is spatial self-similarity in its purest form !
\begin{figure*}[t]
\begin{center}
\vskip -1.0truecm
\mbox{\hskip 0.5truecm\includegraphics[width=16.0cm]{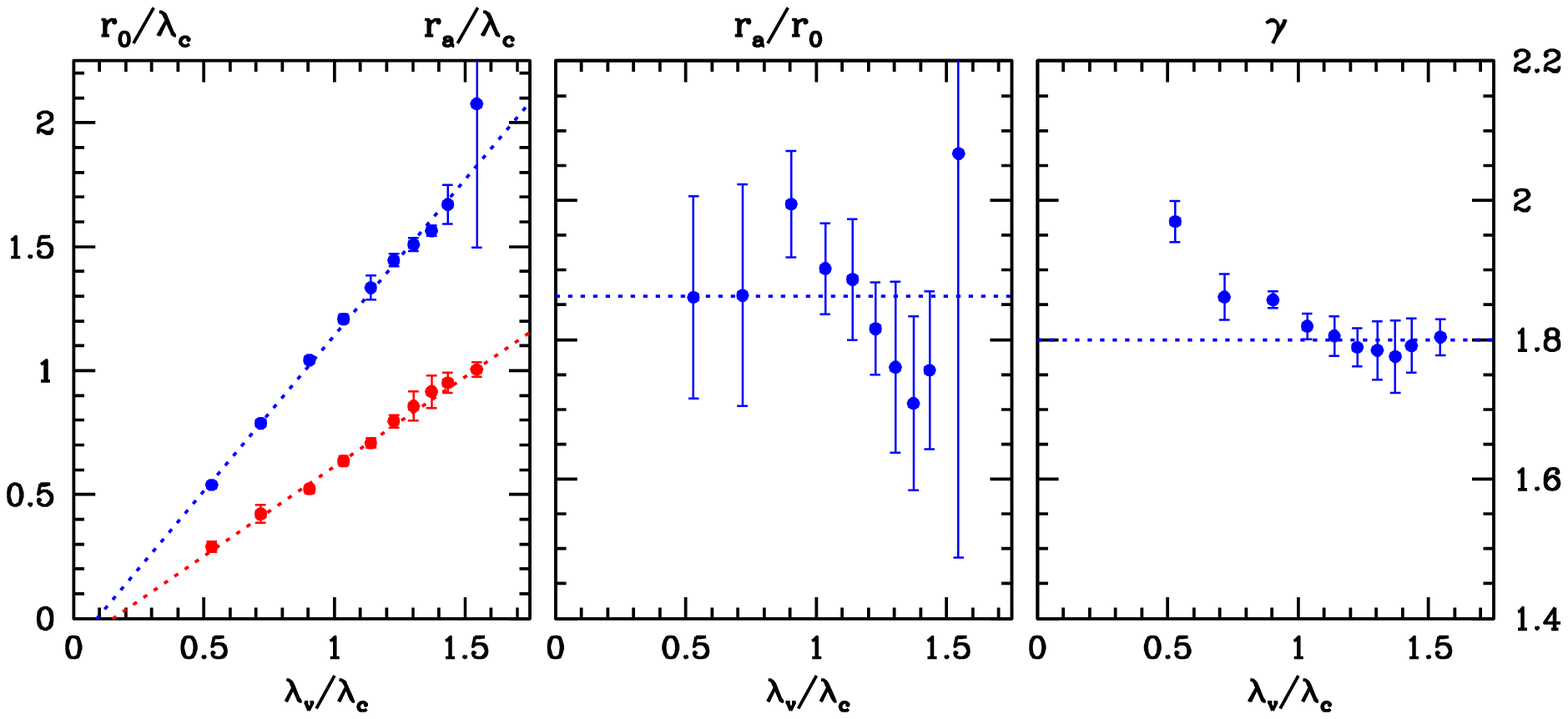}}
\end{center}
\vskip -8.5truecm
\caption{Scaling of Voronoi vertex two-point correlation function parameters, as function 
of average vertex separation in (mass) selected subsample, $\lambda_v/\lambda_c$.   
Left: clustering length $r_0$ (lower, $\xi(r_0) \equiv 1.0)$; coherence length $r_a$ (top, $\xi(r_a) \equiv 0.0)$. 
Centre: ratio clustering-coherence length, $r_a/r_0$. Right: power-law slope $\gamma$.}
\label{fig:corrrscale}
\vskip -0.5truecm
\end{figure*}
\subsection{Vertex Correlation Scaling}
To quantify the impression given by the distribution of the biased vertex selections, 
we analyzed the two-point correlation function for each vertex sample. We computed 
$\xi(r)$ for samples ranging from the complete sample down to the ones merely 
containing the $2.5\%$ most massive ones\footnote{We use the average distance $\lambda_v(R) = 
n(R)^{-1/3}$ between the sample vertices for characterizing the richness of the 
sample. This is based on $\lambda_v$ increasing monotonously with subsample richness.} 

The surprising finding is that all subsamples of Voronoi vertices do retain a two-point 
correlation function (Fig.~\ref{fig:xicorrvrtxscale}) displaying the same qualitative behaviour 
as the $\xi_{\rm vv}(r)$ for the full unbiased vertex sample (Fig.~\ref{fig:vorvrtxi}). Out to a 
certain range it invariably behaves like a power-law (lefthand frame), while beyond that range the 
correlation functions all show the decaying oscillatory behaviour that already has been encountered in the 
case of the full sample. Nonetheless, we can immediately infer significant systematic trends. 

$\bullet$\ {\it Clustering Strength.} The {\it first} observation is that the amplitude of the correlation functions 
increases monotonously with rising vertex sample richness. Expressing the amplitude in terms of the ``clustering length'' 
$r_{\rm o}$ we find a striking and almost perfect linear relation (Fig.~\ref{fig:corrrscale}, lefthand frame). 
The ``fractal'' clustering scaling description of \cite{szsch1985}, according to which the clustering 
scale $r_0$ scales with the average object distance $L(r)$, turns out to be implicit for weblike geometries: 
\begin{equation}
\xi_{\rm cc}(r)=\beta\ (L(r)/r)^\gamma\,;\qquad L(R)=n^{-1/3}\,.
\end{equation}
This finding is in line with studies based on cluster samples of rich clusters or 
selected on the basis of their X-ray emission \cite{pcw1992,bhc2003}. They do indeed show a trend of an increasing 
clustering strength as the clusters are richer ($\approx$ massive). 

$\bullet$\ {\it Spatial Coherence.} A {\it second} significant observation, from the large scale behaviour 
inferred from the lin-lin plot, is that $\xi_{\rm vv}$ extends out to larger and larger distances as the 
sample richness is increasing. The oscillatory behaviour is systematically shifting 
outward for the richer vertex samples, reflecting their more extended clustering patterns. Even though the basic 
cellular pattern has a characteristic scale of only $\lambda_c$, the sample of the $5\%$ richest nodes sets up 
coherent patterns at least 2 to 3 times larger (also cf. fig.~\ref{fig:vorvrtxselfsim}). Pursuing the systematics of 
this geometric clustering, Fig.~\ref{fig:corrrscale} also plots the {\it coherence scale} $r_{\rm a}$. It scales 
almost perfectly linear with average vertex distance $\lambda_v$ !!

These finding are in line with observational evidence that $\xi_{cc}(r)$ extends out considerably larger scales than the galaxy-galaxy 
correlation $\xi_{gg}$, possibly out to $50-100h\hmpc$, even though less evident than the increased strength of 
clustering. The inescapable conclusion is that {\it weblike geometries manage to define coherent structures significantly 
larger than their basic size. They induce clustering in which richer objects not only cluster more strongly, but also 
over an ever larger spatial range.} !!! 

$\bullet$\ {\it Self-Similarity.} Combining the behaviour of clustering scale scale $r_{\rm o}$ and coherence scale 
$r_{\rm a}$ a remarkable {\it self-similar}  scaling behaviour is revealed. The ratio of correlation 
versus clustering length is virtually constant for all vertex samples, $r_{\rm a}/r_{\rm o} \approx 1.86$. 
{\it Weblike geometries may involve intriguing self-similar patterns of clustering}. 

$\bullet$\ {\it Clustering Slope.} A {\it final} interesting detail on the vertex clustering 
scaling behaviour is that a slight and interesting trend in the behaviour of power-law slope. The richer 
samples correspond to a tilting of the slope. Fig.~\ref{fig:corrrscale} shows  
a gradual change from a slope $\gamma \approx 1.95$ for the full sample, to a robust 
$\gamma \approx 1.8$ for the selected samples. Note than this suggestive value is found in the majority of 
observed galaxy and cluster samples !

\section{Summary: Self-Similar Cosmic Geometry}
The revealed systematic trends of vertex clustering have uncovered a hidden 
{\it self-similar} clustering of vertices. It forms a tantalizing indication 
for the existence of self-similar clustering behaviour in spatial 
patterns with a cellular or foamlike morphology. It may hint at an 
intriguing and intimate relationship between the cosmic foamlike geometry 
and various measures of clustering in the Universe.

%------------------------------------------------------------------------- 
\bibliographystyle{latex8}
\bibliography{vorweyisvd07}

\end{document}